\documentclass{sig-alternate}
\CopyrightYear{2014}
\graphicspath{{figures/}}
\DeclareGraphicsExtensions{.pdf}
\usepackage{hyperref}

\begin{document}

%
\conferenceinfo{OMNeT++ Community Summit}{'14 Hamburg Germany}

\title{Towards information-centric WSN simulations}
%
%
%
%
%

\numberofauthors{3} 
%
\author{
%
%
\alignauthor
Gabriel Martins Dias\\
       \affaddr{University Pompeu Fabra}\\
       \affaddr{Barcelona, Spain}\\
       \email{gabriel.martins@upf.edu}
\alignauthor
Boris Bellalta\\
       \affaddr{University Pompeu Fabra}\\
       \affaddr{Barcelona, Spain}\\
       \email{boris.bellalta@upf.edu}
\alignauthor Simon Oechsner\\
       \affaddr{University Pompeu Fabra}\\
       \affaddr{Barcelona, Spain}\\
       \email{simon.oechsner@upf.edu}
}

\date{04 July 2014}

\maketitle
\begin{abstract}

In pursuance of integrating Wireless Sensor Networks (WSNs) with other 
systems, the use of techniques from other fields, such as machine learning and 
information processing, are becoming more common. Therefore, we faced the 
problem of missing network simulations that are not only focused on the packet 
exchange between network elements, but also in the data that is transmitted 
between them. In other words, we needed a tool that evaluated the WSNs on how 
they evolve and react to the environmental changes. To illustrate the benefits 
of having such perspective, we explain the kind of simulation problems that we 
solved in our last work. Moreover, we outline the next steps in the direction of 
creating an extension to support this approach.

\end{abstract}


\section{Introduction}

The use of Wireless Sensor Networks (WSNs) on different types of environments 
has become common in the past years. 
As a consequence of this, monitoring systems that were composed only by 
wireless sensor nodes evolved and are now capable of using data from multiple 
sources of data to produce knowledge (which we will refer to as 
\emph{information}).
In order to better utilize this information and be able to react to 
environmental changes, machine learning and information processing techniques 
started to be incorporated to these systems. 
Therefore, the models that mainly focused on the dynamics of network 
protocols and elements, now may also consider the information that they 
produce and how it is handled. 
Incorporating the support to the information processing in a simulator makes it 
possible to reproduce the information flow and the information processing, as 
well as the effect that such information may have on the network dynamics.
This is specially relevant, considering that simulations are a important tool 
on a model validation task.

At the current state, OMNeT++~\cite{Varga2001} does not focus on these 
features. In order to support the information processing and to simulate the 
decision making process, the simulations should be extended to use the networks 
as modules in their structure. With that, they would be able to run, in 
parallel, multiple WSNs that can operate completely independent of each other or 
exchange the knowledge produced by their sensor nodes and produce a richer 
knowledge about the environment. In the end, the simulations could have 
algorithms to increase the WSNs lifetime, as well as their accuracy, their 
reliability and, consequently, the relevance of the information produced.

The remainder of this paper is structured as follows. In 
Section~\ref{sec:problem}, we describe the need for information-centric 
simulations and illustrate it with an example of a system that intelligently 
uses data from external WSNs to improve the operation of its own WSN. 
Section~\ref{sec:contribution} shows how we handled this problem and ran 
simulations using WSNs that exchanged information. Finally, 
Section~\ref{sec:future} outlines the next steps in the direction of having a 
concrete solution for such kind of problem.

\section{Problem statement}
\label{sec:problem}

As a natural evolution, monitoring systems that were primarily composed only 
by WSNs, now tend to be improved with self-manage capabilities, which turns it 
possible to update their behavior according to the data produced by their 
elements. As part of this process, data from external WSNs can be used to reduce 
the energy consumption and improve the quality of the measurements done by 
internal WSNs, called as inter-WSNs information exchange \cite{6583430, 
Pal2012}. The main idea behind this concept is that the data 
gathered by other WSNs can be exchanged via their gateways and used to infer 
and/or predict values that are not available locally. Based on this inference, 
the systems are able to improve the operation of their wireless sensor nodes 
without losing the quality of the information obtained by the gateway.

For example, based on the information received from external systems, it may be 
possible to infer the data that was going to be measured by a specific node and 
calculate what is the uncertainty level of those predictions according to 
correlated measurements~\cite{Osborne2008}. In case of a low uncertainty level, 
the node's operation can be updated by the gateway in order to save energy and 
increase its lifetime, as there is no need to make the measurement. Otherwise, 
the gateway can actively query some missing data or update the WSN's operation 
by deactivating nodes or changing the time between two measurements.

Although it 
may sound a simple and straight solution, there are two main aspects to be 
observed before applying these updates: the current environmental state, which 
may not allow a node to be turned off, for example; and the costs (e.g., the 
energy consumption in the nodes) to disseminate the updates to the nodes, which 
can vary depending on the current WSN state and its topology. These decisions, 
may have a relevant impact in the WSN evolution in time.

If the simulations focus on the information exchanged by the nodes rather 
than on the packet transmissions, it is possible to model more accurate 
scenarios involving WSNs and the environmental changes around them. Hence, for 
instance, we can consider the interaction between multiple WSNs in runtime. To 
simulate such a scenario, it is required a way to evaluate the quality of the 
measurements done by the sensors and the WSNs performance, which usually depends 
on the quality of the information received from other WSNs and may vary 
depending on how they reacted to the last environmental changes. 

The current workaround for this kind of problem is to split the simulation into 
many parts. For example, the results about the energy consumption obtained in 
OMNeT++ are used as input in Matlab and, after processing the information and 
making decisions in Matlab, new OMNeT++ simulations are run to apply the 
reactions.
Besides the high overhead to run the simulations in sequence, the transition 
from one platform to the other may result in missing information about the 
system's state, given that a new simulation may not contain all the information 
about the environment at the end of the last one.

\section{Contribution}
\label{sec:contribution}

In~\cite{IntelligentlyMartins2014}, we used OMNeT++, INET and MiXiM to simulate 
an architecture for exchanging information between external WSNs. For those 
simulations, we developed a special type of gateway. It is directly connected to 
their respective WSN sink nodes, as well as maintain an overlay connection 
with the other gateways in order to exchange information about their networks 
and measured data. Thus, the gateways run an algorithm that infers values 
based on the information that is available for them at runtime. 
Based on the output of this algorithm, they make decisions about changing the 
operation of the nodes and producing more or less measurements.

Apart from developing the algorithm that handles the data in INET, we did small 
changes in MiXiM to use the WSNs as modules in our simulations. With that change 
in perspective, different types of WSNs can run in parallel and exchange the 
information produced by their nodes in real time. The difference from typical 
simulations is that, in this case, the system evolves in time according to the 
data that is transmitted by the wireless sensor nodes. Thus, the data that is 
produced by the algorithm that infers and/or predicts the measurements affect 
the network topology in the next time-interval, which reflects in the further 
decisions in a cascade effect. As usual, the simulation is able to report the 
required time to query data from specific nodes and to transmit them through the 
network, as well as the energy consumption. Based on these values, the algorithm 
used in the gateway may adapt, change or update its parameters to react to the 
environmental changes and work at the optimal operating level.


\section{Future Work}
\label{sec:future}

In OMNeT++, there are many tools to save the statistics and plot data about 
the wireless sensor nodes, but usually the modules that handle the information 
generated by a WSN are developed in other platforms, such as 
Matlab\footnote{\url{http://www.mathworks.es/products/matlab/}} and 
R\footnote{\url{http://www.r-project.org/}}, or in different 
languages other than C++. Also, some of them are part of a commercial software 
or make use of third-part APIs. There is a missing gap between OMNeT++ and 
those external sources, and standard solution for this problem would facilitate 
the information-centric simulations.

Based on the fact that Matlab includes tools that are useful for machine 
learning and information processing, the next steps to have a 
information-centric extension include creating a way to communicate with other 
platforms to process the information produced by the networks during the 
simulation runtime. With these changes, it would be also possible to use the 
simulation results to make decisions in real time systems.

\section{Acknowledgments}
This work has been partially supported by the Spanish Government under project 
TEC2012-32354 (Plan Nacional I+D),  and by the Catalan Government 
(SGR-2014-1173).

%
\bibliographystyle{unsrt}
\bibliography{bibliography}

\end{document}